\documentclass{aastex}
\usepackage{emulateapj5,epsf,graphicx}

\newcommand{\deblok}{\mbox{de Blok}}
\newcommand{\refand}{\&}

\newcommand{\sct}[1]{\mbox{Section \ref{sct:#1}}}
\newcommand{\fig}[1]{\mbox{Figure \ref{fig:#1}}}
\newcommand{\tbl}[1]{\mbox{Table \ref{tbl:#1}}}

\newcommand{\hi}{{\sc Hi}}

\newcommand{\ngc}[1]{\mbox{NGC #1}}
\newcommand{\n}[1]{\mbox{N#1}}
\newcommand{\lgg}[1]{\mbox{LGG #1}}
\newcommand{\eso}[1]{\mbox{ESO #1}}

\newcommand{\rpm}[1]{\mbox{#1\thinspace m}}
\newcommand{\rpcm}[1]{\mbox{#1\thinspace cm}}
\newcommand{\rpmhz}[1]{\mbox{#1\thinspace MHz}}
\newcommand{\rpmpc}[1]{\mbox{#1\thinspace Mpc}}
\newcommand{\rpkpc}[1]{\mbox{#1\thinspace kpc}}
\newcommand{\rps}[1]{\mbox{#1\thinspace s}}
\newcommand{\rpyr}[1]{\mbox{#1\thinspace yr}}
\newcommand{\rpgyr}[1]{\mbox{#1\thinspace Gyr}}
\newcommand{\rpkms}[1]{\mbox{#1\thinspace km\thinspace s$^{-1}$}}
\newcommand{\rparcmin}[1]{\mbox{#1$^\prime$}}
\newcommand{\rpmjy}[1]{\mbox{#1\thinspace mJy}}
\newcommand{\rpjy}[1]{\mbox{#1\thinspace Jy}}
\newcommand{\rppercent}[1]{\mbox{#1\%}}
\newcommand{\rpjykms}[1]{\rpkms{\rpjy{#1}}}
\newcommand{\rpmsun}[1]{#1\thinspace $M_\odot$}
\newcommand{\rppcmsq}[1]{#1\thinspace cm$^{-2}$}

\newcommand{\eg}{{e.g.}}
\newcommand{\ie}{{i.e.}}
\newcommand{\etal}{{et al.}}
\newcommand{\viz}{viz.}

\slugcomment{To appear in AJ, 2001.}

\shorttitle{{\sc Hi} in NGC 3109 and Antlia}
\shortauthors{Barnes and \deblok}

\begin{document}

\title{On the Neutral Gas Content and Environment of NGC 3109 and
  the Antlia Dwarf Galaxy}

\author{D. G. Barnes}
\affil{Centre for Astrophysics \& Supercomputing, Mail number 31, 
        Swinburne University of Technology, PO Box 218, 
        Hawthorn, VIC 3122, Australia}
\email{dbarnes@swin.edu.au}
\author{W. J. G. de Blok} 
\affil{Australia Telescope National Facility, CSIRO, PO Box 76,
        Epping, NSW 1710, Australia}
\email{edeblok@atnf.csiro.au}

\begin{abstract}
  As part of a continuing survey of nearby galaxies, we have mapped
  the neutral gas content of the low surface brightness,
  Magellanic-type galaxy \ngc{3109} --- and its environment, including
  the Antlia dwarf galaxy --- at unprecedented velocity resolution and
  brightness sensitivity.  The \hi\ mass of \ngc{3109} is measured to
  be \rpmsun{$(3.8 \pm 0.5) \times 10^8$}.  A substantial warp in the
  disk of \ngc{3109} is detected in the \hi\ emission image in the
  form of an extended low surface brightness feature.  We report a
  positive detection in \hi\ of the nearby Antlia dwarf galaxy, and
  measure its total neutral gas mass to be \rpmsun{$(6.8 \pm 1.4)
    \times 10^5$}.  We show the warp in \ngc{3109} to lie at exactly
  the same radial velocity as the gas in the Antlia dwarf galaxy and
  speculate that Antlia disturbed the disk of \ngc{3109} during a mild
  encounter \rpgyr{$\sim 1$} in the past.  \hi\ data for a further
  eight galaxies detected in the background are presented.
\end{abstract}


\section{Introduction}

The morphology and evolution of galaxies are for a major part
determined by their environment.  Spectacular systems such as the
Antennae \citep{arp66, toomre72} and the Cartwheel galaxy
\citep{struckmarcell93, higdon96} are extreme examples of distortion
caused by nearby companions.  Locally, the Magellanic Stream offers a
close-up view of a system dominated by the gravitational field of the
Milky Way Galaxy \citep{putman98}.  Intense infrared emission appears
to be triggered by strong interactions between spirals that are rich
in molecular gas \citep{sanders96}, and there is evidence that low
surface brightness (LSB) galaxies accompanied by nearby dwarf galaxies
will be rapidly transformed into star-bursting blue compact galaxies
because of tidal effects \citep{taylor97}.  N-body simulations support
the observations, showing galaxy harassment to be capable of
transforming disk galaxies into spheroidal galaxies \citep{moore96,
moore98a} and perhaps also of seeding quasar formation in massive, low
surface brightness disks \citep{moore98b}.

Understanding the effects that companions have on their parent
galaxies is therefore important in the context of galaxy evolution.
While a number of galaxies, especially the Milky Way, are known to be
embedded in a halo of dwarf galaxies, very few galaxies are known to
have {\em gravitationally bound}\/ companions.  If, as is often claimed,
galaxy formation is still on-going \citep[\eg][]{wilcots98}, many
galaxies should still be surrounded by a halo of dwarf galaxies and
infalling neutral hydrogen gas (\hi) clouds.  Occasional serendipitous
discoveries have given us tantalising glimpses of these kind of
phenomena, such as infalling dwarf galaxies surrounding Magellanic
irregulars \citep{wilcots96}, and infalling gas clouds in LSB galaxies
\citep{deblok00,deblok99}.  For the systems that are
known, the dynamics of the infalling dwarf galaxies are an excellent
indicator of the presence and mass of dark matter in the parent
galaxy, especially since the dwarfs are found out to much larger radii
than the galaxy disk \citep{zaritsky97}.

To address the lack of information pertaining to companions to known
galaxies, we have commenced a deep \hi\ survey of extended regions
around selected nearby galaxies.  \hi\ observations are ideal for
finding gas-rich dwarf companion galaxies, but large volumes near
target galaxies have hitherto not been surveyed since \hi\
observations that are both wide-field and sensitive have been too
expensive in observing time.  The new Parkes \rpcm{21} Multibeam
receiver \citep{staveleysmith96} however, with its narrowband filters
\citep{haynes98}, offers the opportunity to efficiently image nearby
galaxies and their environments at unprecedented brightness
sensitivity and velocity resolution, with moderate spatial resolution.

To date, we have imaged the \hi\ environments of four galaxies ---
Wolf-Lundmark-Melotte (WLM), \mbox{Sextans A}, \ngc{1313} and
\ngc{3109} --- using this instrument.  Our observations provide lower
image noise, finer velocity resolution and fewer observing artifacts
than standard HIPASS data \citep{barnes01}.  The combined results of
our survey (including further galaxies) will be the subject of a
future paper.  In the meantime, this paper reports the findings
associated with \objectname[]{\ngc{3109}}, a LSB, Magellanic,
late-type spiral.  A brief description of \ngc{3109} and its
environment is given in \sct{n3109}, followed in \sct{obs} by a
description of the observations.  The results of the observations,
pertaining to \ngc{3109}, to background galaxies and to the Antlia
dwarf galaxy, are documented in \sct{results} and discussed in
\sct{discuss}.

\section{\ngc{3109}}
\label{sct:n3109}

\ngc{3109} is a well-studied galaxy at a distance of
\rpmpc{$(1.2\pm0.1)$} \citep{capaccioli92, lee93}, which places it in
the outskirts of the Local Group of galaxies \citep{vandenbergh94}.
The radial velocity of the galaxy is \rpkms{404} \citep{jobin90}.
\ngc{3109} has been variously classified as Irr \citep{sandage61,
  vandenbergh94}, Sm \citep{sandage87} and SB(s)m
\citep{devaucouleurs91}.  In this paper, we consider \ngc{3109} to be
a LSB, Magellanic, late-type galaxy seen edge-on at an inclination of
$75^\circ$ to $80^\circ$ \citep{carignan85, jobin90}.

\ngc{3109} appears to be dark-matter dominated, having a
dark-to-luminous mass ratio of 5--10 \citep{jobin90}.  Stars in the
disk of \ngc{3109} have ages as young as \rpyr{$10^7$}, and a halo of
\mbox{Population II} stars with ages exceeding \rpyr{$10^{10}$} has
been observed \citep{minniti99}.  This is consistent with models of
LSB galaxies which exhibit star formation rates that are on average
low but erratic \cite[\eg][]{gerritsen99}.  Since these galaxies
evolve slowly, the surrounds of \ngc{3109} may be relatively rich in
agglomerations of \hi\ whose densities have not reached those required
for star formation, and which have not yet settled into the disk of
the galaxy.

Already, one possible companion to \ngc{3109} is known: the
\objectname[]{Antlia dwarf galaxy} (hereafter, Antlia), at a distance
of \rpmpc{$(1.3\pm0.1)$} from the Milky Way \citep{sarajedini97,
  aparicio97}.  Like \ngc{3109}, Antlia contains stellar populations
of all ages \citep{sarajedini97}, with most recent star formation
taking place slowly in the central regions \citep{aparicio97}.  In
this respect, it is typical of the dwarf spheroidals in the Local
Group (\eg\ Tucana, Carina).  The radial velocity of Antlia is
\rpkms{361} \citep{fouque90}.  \cite{aparicio97} calculate the
physical separation of \ngc{3109} and Antlia to be between \rpkpc{29}
and \rpkpc{180}, with a maximum separation of \rpkpc{37} for the pair
to be gravitationally bound.

\section{Observations}
\label{sct:obs}

We have acquired a deep \hi\ image of a field \rpkpc{120} square
centered on \ngc{3109} and including Antlia.  Our observations
utilise the new \rpcm{21} Multibeam system at the prime focus of the
Parkes \rpm{64} radiotelescope.\footnote{The Parkes radiotelescope is
part of the Australia Telescope which is funded by the Commonwealth of
Australia for operation as a National Facility managed by CSIRO.}
With narrowband filters installed, the instantaneous observing
bandwidth is \rpmhz{8}, which is correlated into 2048 channels for two
products XX and YY.  We centered the observing band at \rpmhz{1418.4},
corresponding to a radial velocity of \rpkms{424}, with channels
extending \rpkms{844} either side in increments of \rpkms{0.83}.

To image \ngc{3109} and its surroundings, scans at a rate of
\mbox{$1^\circ$\thinspace min$^{-1}$} were made separately in the
Declination and Right Ascension directions, across a square field of
side length $5.67^\circ$.  Adjacent scans were separated by a maximum
of \rparcmin{4}, so that each beam of the Multibeam system sampled the
\hi\ sky at better than the Nyquist rate of the beam.  A total of 208
scans were made, comprising $\sim 200000$ single-polarisation spectra.
The individual spectra, recorded every \rps{5}, were
bandpass-corrected, calibrated and imaged using the software developed
for the \hi\ Parkes All Sky Survey \citep[HIPASS,][]{barnes01}, with
the HIPASS parameters slightly modified to accommodate the shorter
scan length of this survey.  The final image, having pixels of side
length \rparcmin{4} spatially and \rpkms{$0.83$} spectrally, had an RMS
noise level of \rpmjy{34}.  The image beam (resolution) is
\rparcmin{$\sim15.5$} FWHM spatially, and \rpkms{1.12} FWHM
spectrally.

\section{Results}
\label{sct:results}

\subsection{The \hi\ Content of \ngc{3109}}

\begin{figure*}
\includegraphics[width=0.7\hsize]{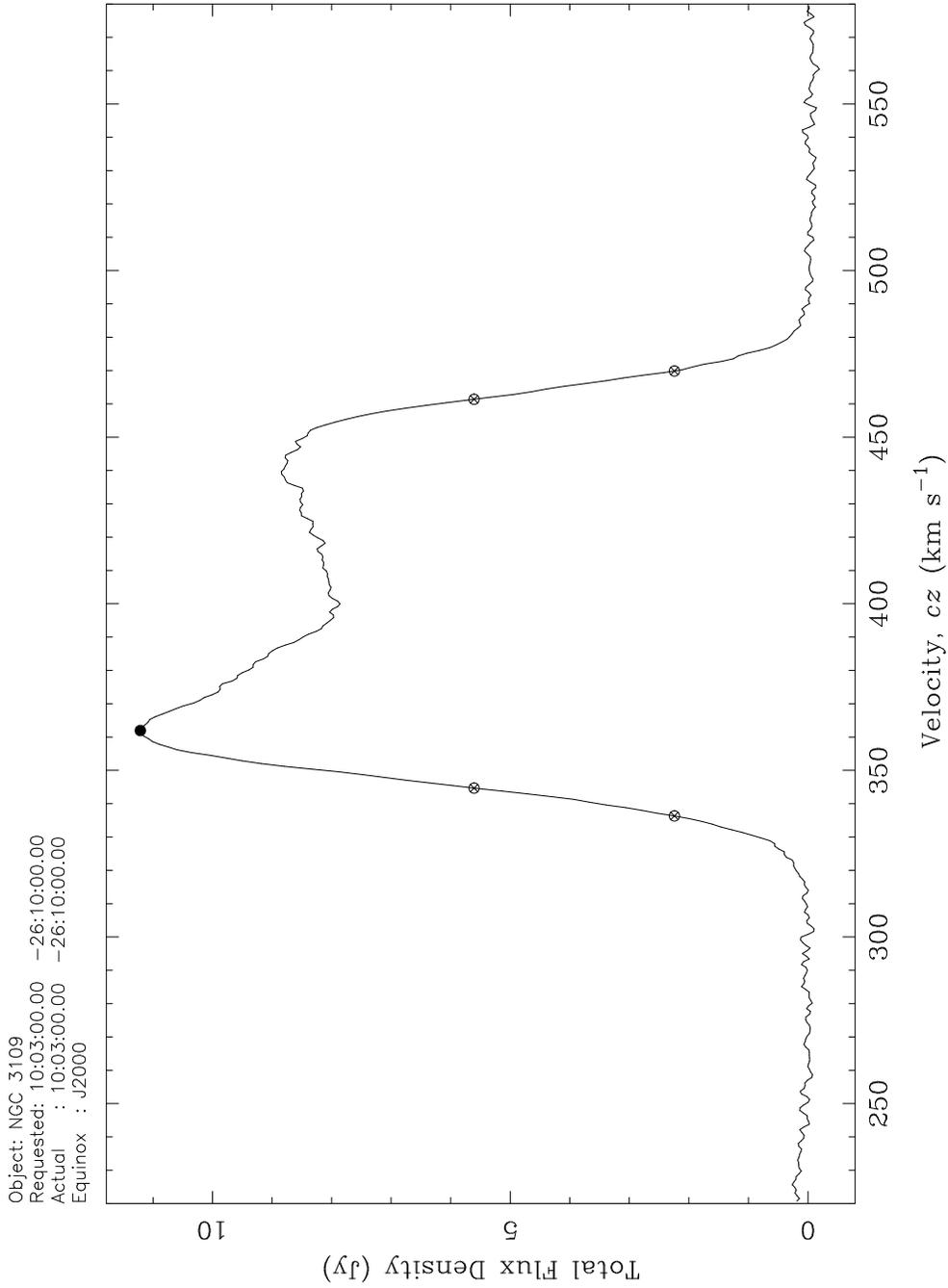}
\caption{The integrated \hi\ profile of \ngc{3109},
  smoothed with a Hanning filter of width three channels.  The
\rppercent{50} and \rppercent{20} peak flux points are marked
(width-minimised points with crosses, width-maximised with hollow
circles), along with the peak flux itself (solid circle) at
\rpkms{362}.\label{fig:n3109spec}}
\end{figure*}

The spectrum of \hi\ emission integrated spatially over \ngc{3109} is
shown in \fig{n3109spec}.  The overall profile shape is markedly
asymmetric, having a surplus of flux in the lower velocity component
of the disk.  This property of \ngc{3109} has already been noted by
\cite{huchtmeier80}, and will be discussed below.  The line centre is
located at \rpkms{$(403.0 \pm 0.2)$}, being the coincident midpoint of
the \rppercent{20} and \rppercent{50} peak flux points on the
edges of the profile.  This is in excellent agreement with all
previously published values, \eg\ \rpkms{$(404 \pm 2)$} measured by
\cite{jobin90} and \rpkms{$(403 \pm 2)$} measured by
\cite{huchtmeier73}.  The flux-weighted mean velocity is
\rpkms{$(399.5 \pm 0.5)$}, and the peak flux is detected at
\rpkms{362}.  The peak projected rotation velocity of the galaxy is
\rpkms{$(72.4 \pm 0.5)$}.

The integrated \hi\ flux of \ngc{3109}, measured by the traditional
zeroth-order moment of the spectrum, and verified with an equivalent
but robust statistic, is \rpjykms{$1110_{-30}^{+90}$}; the uncertainty
in this value is dominated by the uncertainty in the image beam
\citep[see][]{barnes01}.  This value compares well to \rpjykms{1150}
measured from standard HIPASS data, and is also recovered when the
narrowband spectral data is re-imaged in the traditional way (\ie\ 
with a Gaussian smoothing kernel and using the mean estimator instead
of the median).  Compared to the spectrum published by \cite{jobin90},
from which an integrated \hi\ flux of \rpjykms{$(650 \pm 130)$} is
measured, we detect substantially more flux in \ngc{3109} at all
velocities.  This is expected, since \cite{jobin90} used the Very
Large Array (VLA) in a configuration blind to emission on scales much
greater than \rparcmin{$\sim20$} on the sky.  Of the many single dish
studies made of \ngc{3109}, the flux measured here is consistent only
with the value \rpjykms{$(1280 \pm 150)$}, measured by
\cite{epstein64} with the Harvard 60-ft antenna, whose beam is
sufficiently large to ``see'' all of \ngc{3109}.  Other single dish
measurements, \viz\ \rpjykms{530} \citep{whiteoak77}, \rpjykms{1390}
\citep{dean75}, \rpjykms{1460} \citep{huchtmeier73} and \rpjykms{1660}
\citep{huchtmeier80}, were made using larger telescopes, and we
suggest that these measurements have been compromised by large scale
insensitivity. 

We deduce an \hi\ mass for \ngc{3109} of \rpmsun{$(3.8 \pm 0.5) \times
  10^8$}, not corrected for \hi\ line opacity.  This is lower than
most previously determined values for the following reasons: there has
been a general trend towards lower distance estimates to \ngc{3109} in
the more recent literature; the majority of previous \hi\ flux
measurements seem too high as described above; no correction of \hi\ 
optical depth has been made in this work.  The synthesis \hi\ mass of
\cite{jobin90}, corrected for the recent (lower) distance measures to
\ngc{3109}, is \rpmsun{$(2.2 \pm 0.5) \times 10^8$}.  Thus the \hi\ 
mass not detected by the VLA observations is of order
\rpmsun{$1.6 \times 10^8$}.

\begin{figure*}
\includegraphics[width=0.4\hsize]{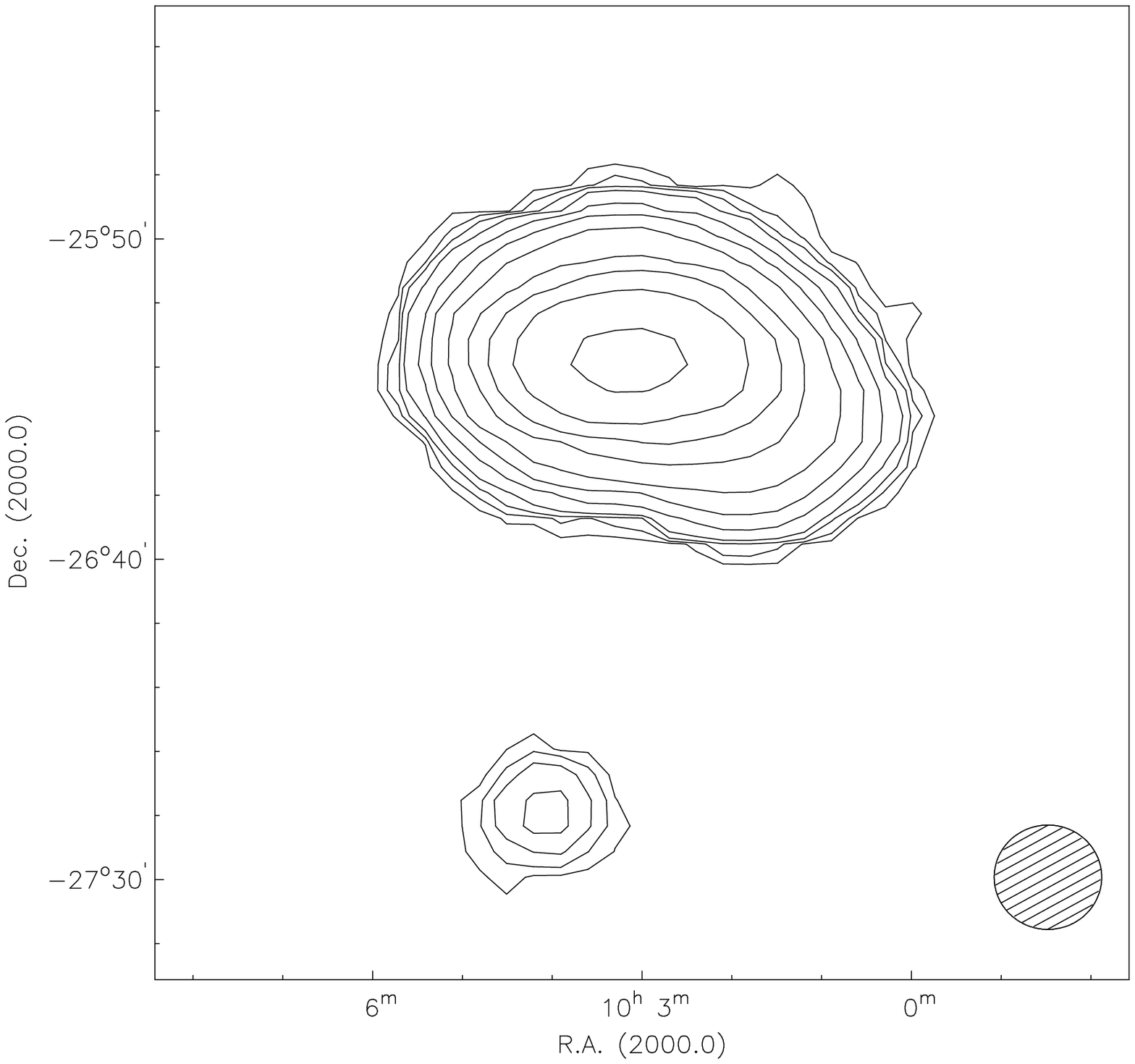}
\includegraphics[width=0.4\hsize]{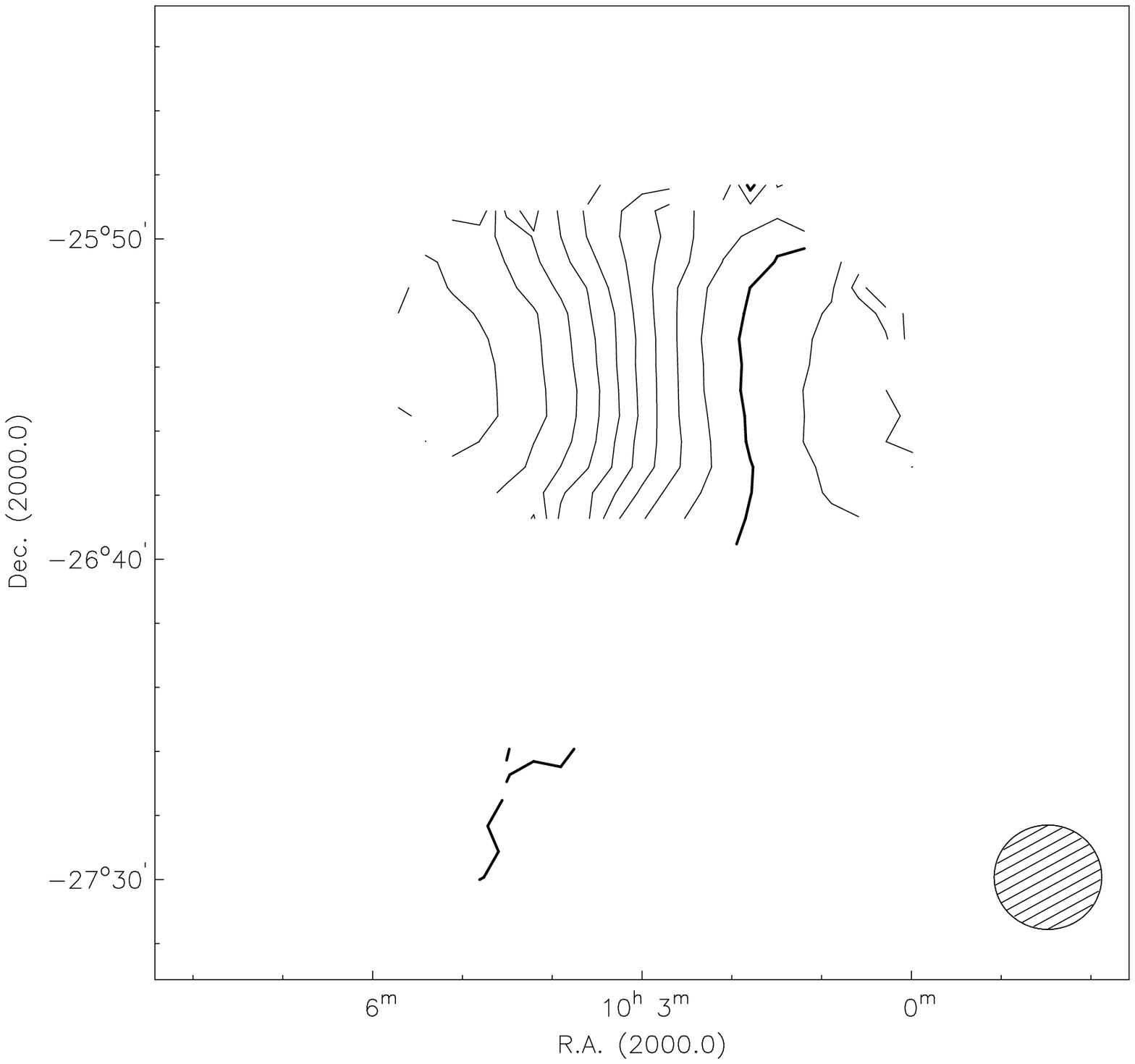}
\caption{Integrated \hi\ column density map
  (left) and velocity field (right) of \ngc{3109} and the Antlia dwarf
  galaxy.  The column density contours are placed at 2, 5, 10, 20, 50,
  100, 200, 500, 1000, 2000 and \mbox{$5000 \times 10^{17}$
    cm$^{-2}$}.  The velocity contours increase by \rpkms{10}
  eastward, and the bold contour lies at \rpkms{360}.  The approximate
  observing beam size is given at the lower right of each
  map.\label{fig:moments}}
\end{figure*}

In \fig{moments}, integrated \hi\ column density and flux-weighted
velocity maps of \ngc{3109} are given.  A warp in the column density
and velocity field is evident in the south west of the galaxy.
Inspection of individual channel maps demonstrates that this component
is the source of the asymmetry in the (single dish) integrated \hi\ 
profile of the galaxy, and that it contains a substantial fraction of
the neutral gas not seen by the VLA, in the form of a smooth, extended
component near velocity \rpkms{360}.  We note that this reservoir of
gas lies in that part of velocity space which coincides with the
radial velocity of Antlia.  This will be discussed further below.

\subsection{Background galaxies}

The field imaged by our observations measures \rpkpc{120} square at a
distance of \rpmpc{$1.2$}.  As expected, the Antlia dwarf galaxy was
detected, and its \hi\ parameters are presented below.  Besides
\ngc{3109} and Antlia, eight previously catalogued galaxies were
detected in the field, at radial velocities between \rpkms{$\sim 850$}
and \rpkms{1150}.  Data pertaining to these galaxies --- including the
\hi\ properties measured in this survey --- are given in
\tbl{detections}.  Two of the galaxies --- \ngc{3113} and \eso{499-
  G037} --- have previously been identified as members of the galaxy
group \lgg{189} \citep{garcia93}.  The radial velocities of \lgg{189}
and of the remaining \hi\ galaxies in the field range places them at
least \rpmpc{$\sim 12~h_{75}^{-1}$} behind \ngc{3109} and Antlia,
allowing us to conclude that it is highly unlikely that they have
exerted any influence on the formation or evolution of \ngc{3109}.

\subsection{The \hi\ Content of the Antlia Dwarf Galaxy}

\begin{figure*}
\includegraphics[width=0.7\hsize]{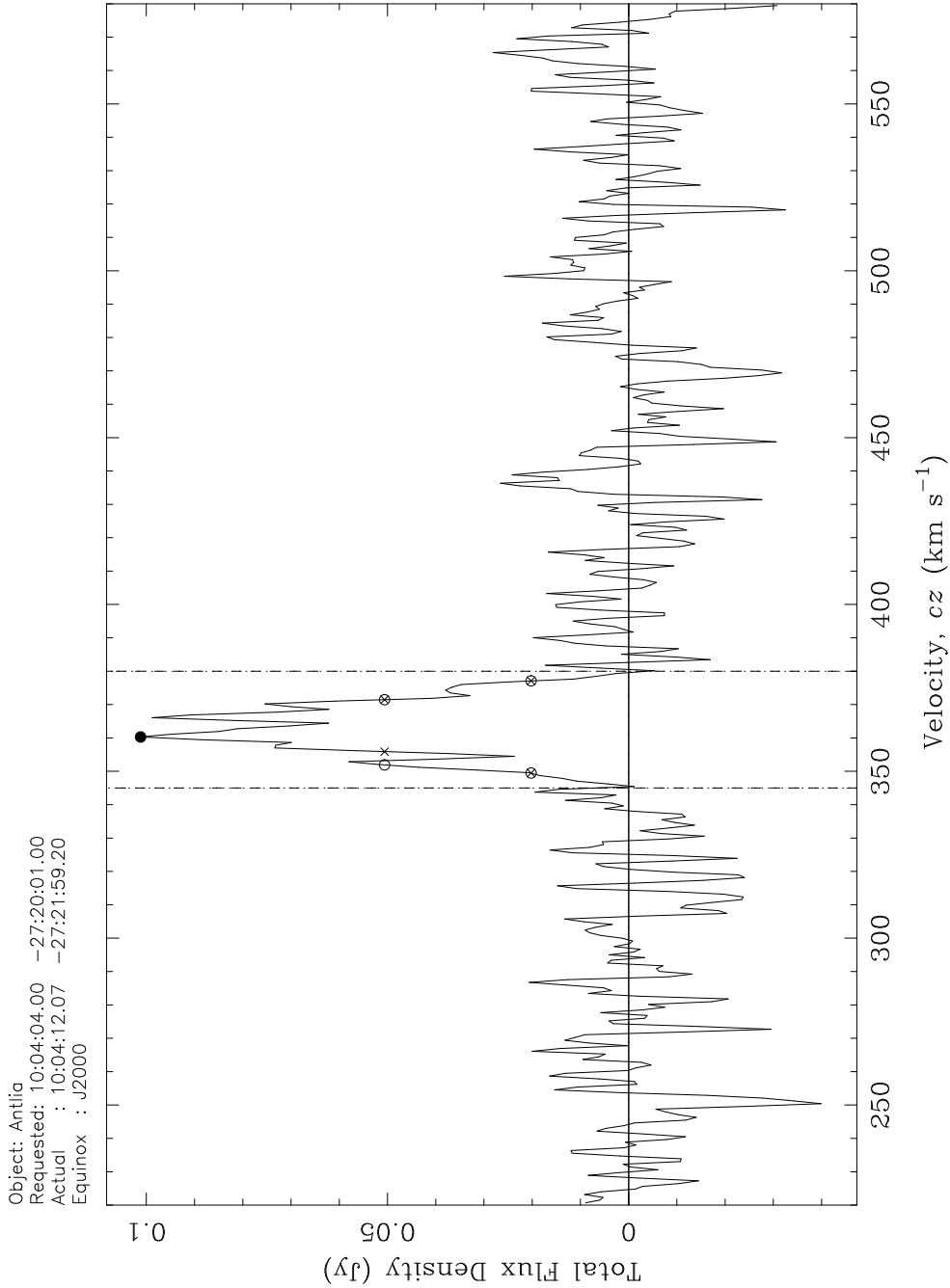}
\caption{An \hi\ spectrum taken at the position of the
Antlia dwarf galaxy, integrated over an area \rparcmin{20} square, and
smoothed with a Hanning filter three channels wide.  The baseline
was removed with a seventh order fit to the channels exterior to the
region marked by the dashed lines.  The \rppercent{50} and
\rppercent{20} peak flux points are marked as for \fig{n3109spec}, as
is the peak flux itself at \rpkms{360}. \label{fig:antliaspec}}
\end{figure*}

The \hi\ emission spectrum extracted at the position of the Antlia
dwarf galaxy and integrated over a region $20^\prime \times
20^\prime$, is shown in \fig{antliaspec}.  The galaxy is easily
detected, and is well distinguished from \ngc{3109} emission.  The
spectrum is that of a single peak profile with a maximum flux of
\rpmjy{100} at \rpkms{360}.  The peak flux measured here is consistent
with the previously determined upper limits of \rpmjy{150}
\citep{gallagher95} and \rpmjy{122} \citep{fouque90}.  The line centre
is measured to be \rpkms{$(362.0 \pm 2.0)$}, this being the mean of
the midpoints of the \rppercent{20} and \rppercent{50} peak flux
points on the edges of the profile, in exact agreement with the
detection by \cite{fouque90}, and coinciding with the flux-weighted
mean velocity.  The velocity width is \rpkms{$(30 \pm 2)$}.  We note
that the detection by \cite{fouque90} is almost certainly real,
contrary to the suggestion of \cite{blitz00} that it is an
instrumental artifact.

Summing the flux in the \hi\ profile of Antlia yields a total flux of
\rpjykms{$(1.7 \pm 0.1)$}.  Under the usual assumptions, and adopting
a distance of \rpmpc{$(1.3 \pm 0.1)$} to Antlia, this yields an \hi\ 
mass of \rpmsun{$(6.8 \pm 1.4) \times 10^5$}.  Our result is
considerably lower than that of \cite{whiting97}, who used the data of
\cite{fouque90} to determine a total \hi\ mass of Antlia of order
\rpmsun{$(1.0 \pm 0.2) \times 10^6$},\footnote{We have corrected their
  result for our adopted distance to Antlia.} and lies within the
range of \hi\ contents detected in dwarf spheroidal galaxies by
\cite{blitz00}.  Moment analysis applied to the \hi\ spectrum of
Antlia yields a line-of-sight RMS velocity of \rpkms{$6.4\pm0.7$}.
Following \cite{whiting97}, a central mass-to-light ratio of
$(7.6\pm1.6) M_\odot / L_\odot$ is calculated for Antlia.

\section{Discussion}
\label{sct:discuss}

\subsection{Sensitivity}

For a point source at the distance of \ngc{3109}, the minimum
detectable \hi\ mass (3$\sigma$ per channel) is of order \rpmsun{$4
  \times 10^4 \Delta V$} where $\Delta V$ is the velocity width of the
galaxy expressed in \rpkms{}.  We have neglected the improvement in
detectability as source width increases, and so the Antlia dwarf
galaxy, which extends over more than 30 channels, actually lies
slightly below this limit.  For a smooth distribution of \hi\ filling
the beam, the minimum detectable column density for these observations
is \rppcmsq{$10^{17}$} (3$\sigma$, per \rpkms{0.83} channel).  Such a
gas cloud would have a neutral gas mass similar to that calculated for
the point source case.

\subsection{Dwarf galaxy populations}

In the cube of space centred on \ngc{3109}, with sidelength
\rpkpc{120}, we have detected only one dwarf galaxy -- Antlia.  No
other isolated reservoirs of neutral gas were found, and so at this
point little can be deduced about the dark matter content of
\ngc{3109} and its surrounds.  To escape
detection, a gas cloud in the vicinity of \ngc{3109} could be:
\begin{enumerate}
\item{spatially extended but have a neutral gas column density less 
than \rppcmsq{$\sim 10^{17}$},}
\item{smaller in spatial extent than \rpkpc{$\sim 4$} and less massive
than \rpmsun{$\sim 4 \times 10^4\Delta V$}, or be} 
\item{unresolved from \ngc{3109}.}
\end{enumerate}
Objects in the first category are exceedingly difficult to detect
without committing substantial telescope time, and even then will
likely be invisible in \hi\ emission given the typical column
density cutoff seen in galaxy disks at \rppcmsq{$2 \times 10^{19}$}
\citep{corbelli93}.  Nevertheless, such objects must surely exist as
a component of the Ly alpha absorber population.

Dwarf galaxies bound to \ngc{3109} will primarily fall into the second
and third categories.  Further \hi\ imaging at higher spatial
resolution can be used to search for gas clouds unresolved from
\ngc{3109} (category three), but the Parkes \rpcm{21} Multibeam
remains the best instrument for finding the low \hi\ mass
objects of category two.

Our findings are consistent with other recent searches for extremely
low \hi\ mass objects.  For example, the Arecibo \hi\ strip survey of
\cite{zwaan00} yielded an \hi\ mass function predicting a number
density of \mbox{$\sim2$ Mpc$^{-3}$} for objects having \hi\ masses in
the decade \rpmsun{$10^{4.5}$ -- $10^{5.5}$}.  This gives a detection
probability of \mbox{$\sim0.5$ per cent} for a single object in this
mass range in the present study of \mbox{$2 \times 10^-3$ Mpc$^3$}.

\subsection{\ngc{3109} and Antlia}

The warp of the gas disk of \ngc{3109} detected here is also suggested
in the data of \cite{jobin90}.  Their \mbox{Figure 4} shows a narrow
plume of low column density gas extending \rpkpc{$\sim 5$} southwest
from the disk.  Our observations suggest that the gas plume is sitting
on top of a more massive but smooth distribution of \hi\ which is
invisible to the synthesis observations. \cite{jobin90} note that the
\hi\ plume, together with the slight southwest displacement of the
optical isophotes, is suggestive of a past interaction with another
system, yet are unable to suggest a suitable nearby system.  Since the
plume (or warp) lies at exactly the radial velocity of Antlia, we
speculate that Antlia is this very companion, and that some mild
interaction has occurred between \ngc{3109} and Antlia in the distant
past.

For a current separation of \rpkpc{$\sim 50$}, and a differential
velocity of \rpkms{$\sim 50$}, an interaction may have taken place
\rpgyr{$\sim 1$} ago.  The encounter would need to have been prolonged
enough to draw out a substantial mass of neutral gas (of order
\rpmsun{$10^8$}) but mild so as not to cause massive star formation --
the plume is presently undetected optically.  Recent n-body
simulations of the Milky Way and Magellanic Clouds
\citep[\eg][]{yoshizawa98} have shown that such encounters can occur.
Further N-body simuiations of the \ngc{3109} and Antlia system are
needed to assess whether this proposed past encounter is feasible.

\acknowledgments

We are grateful to the staff of the Parkes observatory -- in
particular John Reynolds and Mal Smith -- for their expert assistance
during the observations.  We are also grateful for the continuing
support of the {\sc livedata} software by Mark Calabretta and Lister
Staveley-Smith of the Australia Telescope National Facility.





\clearpage



\clearpage

\begin{deluxetable}{lccccccc}
\tabletypesize{\scriptsize}
\tablecaption{\hi\ detections in the \ngc{3109} field. \label{tbl:detections}}
\tablewidth{0pt}
\tablehead{
\colhead{Source} & \colhead{R.A.} & \colhead{Dec} & \colhead{Mag} &
\colhead{Line centre} & \colhead{Line width} &
\colhead{\hi\ flux} & \colhead{\hi\ mass} \\
& \multicolumn{2}{c}{(J2000)} & (Cousins B$_T$) & 
\multicolumn{2}{c}{(km s$^{-1}$)} & (Jy km s$^{-1}$) & ($M_\odot$) 
}
\startdata
\eso{499- G010} (\n{3037}) & 09:51:24 & $-$27:00:36 & 13.7 & \phn$877\pm3$   & \phn$96\pm3$ & \phn\phn\phn7.0\phd$^{+\phn0.6}_{-\phn0.2}$                   & $(2.3 \pm 0.3) \times 10^8$ \\
\eso{435- G007} (\n{3056}) & 09:54:33 & $-$28:17:55 & 12.6 & \phn$969\pm1$   &    $272\pm2$ &    \phn\phn46\phd\phn\phd$^{+\phn4\phd\phn}_{-\phn1\phd\phn}$ & $(1.8 \pm 0.2) \times 10^9$ \\
\eso{435- G016}              & 09:58:47 & $-$28:37:22 & 13.4 & \phn$975\pm1$ &    $150\pm1$ &    \phn\phn33.2\phd$^{+\phn2.7}_{-\phn0.9}$                   & $(1.3 \pm 0.2) \times 10^9$ \\
\eso{435-IG020}              & 09:59:21 & $-$28:07:54 & 14.4 & \phn$978\pm3$ &    $112\pm4$ & \phn\phn\phn7.8\phd$^{+\phn0.6}_{-\phn0.2}$                   & $(3.1 \pm 0.4) \times 10^8$ \\
\eso{499- G036} (\n{3109}) & 10:03:07 & $-$26:09:32 & 10.3 & \phn$403\pm1$   &    $144\pm1$ &        1110\phd\phn\phd$^{+90\phd\phn}_{-30\phd\phn}$         & $(3.8 \pm 0.5) \times 10^8$ \\
\eso{499- G037}              & 10:03:42 & $-$27:01:40 & 13.3 & \phn$955\pm1$ &    $212\pm1$ &    \phn\phn49\phd\phn\phd$^{+\phn4\phd\phn}_{-\phn1\phd\phn}$ & $(1.9 \pm 0.3) \times 10^9$ \\
\eso{499- G038}              & 10:03:50 & $-$26:36:46 & 15.7 & \phn$888\pm3$ &    $104\pm2$ &    \phn\phn11.4\phd$^{+\phn1.0}_{-\phn0.3}$                   & $(3.8 \pm 0.5) \times 10^8$ \\
Antlia Dwarf Galaxy          & 10:04:04 & $-$27:20:01 & --   & \phn$362\pm2$ & \phn$30\pm2$ & \phn\phn\phn1.7\phd$^{+\phn0.2}_{-\phn0.1}$                   & $(6.8 \pm 1.4) \times 10^5$ \\
\eso{435- G035} (\n{3113}) & 10:04:26 & $-$28:26:35 & 13.4 &    $1082\pm1$   &    $226\pm1$ &  \phn\phn40\phd\phn\phd$^{+\phn4\phd\phn}_{-\phn1\phd\phn}$   & $(2.0 \pm 0.3) \times 10^9$ \\
\eso{435- G047}              & 10:09:07 & $-$29:03:51 & 12.2 &    $1105\pm1$ &    $268\pm1$ &       \phn154\phd\phn\phd$^{+12\phd\phn}_{-\phn4\phd\phn}$    & $(7.9 \pm 1.0) \times 10^9$ \\
\enddata



\tablecomments{For \n{3109} and Antlia, \hi\ masses have been
  determined using distances as cited in \sct{n3109}; for all other
  galaxies mass calculations Hubble distances have been used with
  \mbox{$H_0 = 75$ km s$^{-1}$ Mpc$^{-1}$}.}

\end{deluxetable}

\end{document}